\documentclass[prb,twocolumn,showpacs,preprintnumbers,amsmath,amssymb]{revtex4}
\usepackage{graphicx}
\usepackage{dcolumn}
\usepackage{amssymb}
\usepackage{amsmath}
\usepackage{verbatim}

\begin{document}

\title{Band-edge problem in the theoretical determination of defect energy levels: the O vacancy in ZnO as a benchmark case}

\author{Audrius Alkauskas$^{1,2}$}
\author{Alfredo Pasquarello$^{2}$}
\affiliation{$^1$Institute of Condensed Matter Physics, Ecole Polytechnique F\'ed\'erale de Lausanne (EPFL), CH-1015 Lausanne, Switzerland\\
$^2$Chaire de Simulation \`{a} l'Echelle Atomique (CSEA), Ecole Polytechnique F\'ed\'erale de Lausanne (EPFL), CH-1015 Lausanne, Switzerland
}

\date{\today}

\begin{abstract}
Calculations of formation energies and charge transition levels of defects routinely rely on
density functional theory (DFT) for describing the electronic structure. Since bulk
band gaps of semiconductors and insulators are not well described in semilocal
approximations to DFT, band-gap correction schemes or advanced theoretical models 
which properly describe band gaps need to be employed. However, it has 
become apparent that different methods that reproduce the experimental band gap
can yield substantially different results regarding charge transition levels of point defects.
We investigate this problem in the case of the (+2/0) charge transition level 
of the O vacancy in ZnO, which has attracted considerable attention 
as a benchmark case. For this purpose, we first perform calculations based 
on non-screened hybrid density functionals, and then compare our results with 
those of other methods. While our results agree very well with those obtained 
with screened hybrid functionals, they are strikingly different compared to those 
obtained with other band-gap-corrected schemes. Nevertheless, we show that all 
the different methods agree well with each other and with our calculations 
when a suitable alignment procedure is adopted. The proposed procedure 
consists in aligning the electron band structure through an external potential,
such as the vacuum level. When the electron densities are well reproduced, this 
procedure is equivalent to an alignment through the average electrostatic 
potential in a calculation subject to periodic boundary conditions.
We stress that, in order to give accurate defect levels, a theoretical scheme 
is required to yield not only {\it band gaps} in agreement with experiment, but 
also {\it band edges} correctly positioned with respect to such a reference potential.
\end{abstract}
\pacs{
      71.15.Nc,   
      71.55.-i,   
      71.55.Gs    
}
\maketitle

\section{Introduction}

Point defects can affect the properties of solids in a dramatic way.\cite{Stoneham} They determine, for example, 
the conductivity of semiconductors, the color of natural crystals, and the mechanical properties of materials. 
Equally important, defects influence or govern the performance and the long-term stability of a wide range of 
semiconductor devices, such as metal-oxide-semiconductor field-effect transistors, photovoltaic cells, solid fuel 
cells, to name a few. The theoretical characterization of defects, especially in wide band-gap materials, 
has become increasingly important in the attempt to understand and control the performance
of these devices.\cite{VanDeWalle_JAP_2004,Wiley} In the last decades, density functional theory (DFT) has grown 
into the standard theoretical model to describe the electronic and atomic structure of solids. The common 
approximations to DFT, \emph{viz.}\ the local density approximation (LDA) and the generalized gradient 
approximation (GGA), systematically underestimate band gaps of semiconductors and insulators.
Since the band gap is the relevant energy scale in the study of defects, this so-called ``band-gap problem'' of
LDA and GGA severely affects the predictive power of these approximations when applied to defect levels.
Recently there have been lots of efforts to assess the importance of band gap corrections
\cite{VanDeWalle_JAP_2004,Wiley,Lany_PRB_2008,Alkauskas_PRL_2008a,Lany_MSMSE_2009,Lambrecht_pssb_2010} and to
use theoretical models giving a much more appropriate description of the bulk band structure.
The choice of methods is large and includes the LDA+$U$ method,\cite{Janotti_APL_2005,Lany_PRL_2007,
Janotti_PRB_2007,Paudel_PRB_2008} approximate self-interaction correction schemes,\cite{Pemmaraju_PRB_2008} 
hybrid density functionals, \cite{Deak_JPCM_2005,Knaup_PRB_2005,Gavartin_PRB_2006,Broqvist_APL_2006,Oba_PRB_2008,
Janotti_PRB_2010,Deak_PRB_2010,Broqvist_pssa_2010,Clark_PRB_2010} the use of modified 
pseudopotentials,\cite{Segev_PRB_2007} empirical schemes,\cite{Persson_PRB_2005} and more advanced theoretical tools, 
such as the many-body perturbation theory within the $GW$ and higher approximations.
\cite{Hedstrom_PRL_2006,Rinke_PRL_2009,Lany_PRB_2010a,Bockstedte_PRL_2010,Bruneval_PRB_2011,Giantomassi_pssb_2011} 


It appears evident to assume that a good theoretical model must at least satisfy two conditions, namely
{\it (i)} give an accurate electron density of the defect system and {\it (ii)} yield a good band gap of the host material. 
While these two requirements form a necessary prerequisite to obtain reliable results concerning
defect formation energies and associated charge transition levels,\cite{VanDeWalle_JAP_2004,Lany_PRB_2008}
it has recently become apparent that it is by no means sufficient. This is best exemplified in the case of defect 
energy levels in ZnO.\cite{Lany_PRB_2008,Janotti_APL_2005,Lany_PRL_2007,Janotti_PRB_2007,Paudel_PRB_2008,Oba_PRB_2008,
Clark_PRB_2010,Kohan_PRB_2000,Erhardt_PRB_2006,Agoston_PRL_2009,Gallino_JCP_2010,Boonchun_pssb_2011} 
This is a particularly severe case, because the LDA and the GGA yield a bulk band-gap of 0.6-0.8 eV, severely
underestimating the experimental value of 3.44 eV. 
For the case of the (+2/0) charge transition level of the oxygen vacancy (V$_{\text{O}}$) 
theoretical models yield levels either as low as 0.6 eV above the valence band maximum (VBM) or as 
high as 2.4 eV above VBM. These results differ significantly  despite the fact that in all these theoretical models  
the ``band-gap problem'' was accounted for. In addition, other critical issues, such as  
finite-size effects associated to the supercell treatment, were presumably under control in these studies.  
Furthermore, the first condition concerning the accuracy of the electron density was clearly also fulfilled 
since the involved electronic state corresponds to the fully symmetric $a_1$ state which is already correctly 
described via a semilocal functional. The second condition concerning the band gap was fulfilled by construction.

Recently, Lany and Zunger provided a very detailed overview of the way various theoretical
and computational approximations affect the determination of defect formation energies and 
charge transition levels.\cite{Lany_PRB_2008} They concluded that, in addition to the two requirements discussed above, 
a reliable theoretical model should correctly describe the relative positions of all relevant 
electronic states. For ZnO, this condition mainly concerns the position of the 
Zn 3$d$ states with respect to the conduction and valence band edges. The importance of this 
requirement becomes evident when considering shallow defects, the wavefunctions of which 
can be always thought as arising from a linear combination of bulk bands.   

In this work, we show that that there is yet another crucial requirement that the theoretical model
must fulfill. In order to yield an appropriate description of defect formation energies and associated charge transition 
levels, the positions of the VBM and the conduction band minimum (CBM) with respect to a suitably defined reference potential should also be 
accurately described. To demonstrate this, we first calculate the (+2/0) charge transition level 
of the V$_{\text{O}}$ in ZnO and compare our result with those available in the literature. 
Our study adds to a series of studies,\cite{Lany_PRL_2007,Lany_PRB_2008,Janotti_PRB_2007,Paudel_PRB_2008,Agoston_PRL_2009,Lany_PRB_2010a,Boonchun_pssb_2011} 
in which conflicting results were found. However, we show that these seemingly incompatible
findings agree reasonably with each other when an alternative alignment scheme is used. 
We provide theoretical arguments to rationalize this finding.  Similar results are expected for other atomically localized 
defects and for other materials in which the ``band-gap problem'' of semilocal calculations is particularly severe. 
Our investigation thus leads to a deeper understanding of the ``band-edge problem'' in the theoretical study of 
defect levels and provides a requirement for the theoretical model in addition to the conditions mentioned above.

This paper is organized as follows. In Sec.\ \ref{Comp}, we summarize our computational approach for calculating 
defect formation energies and charge transition levels. The obtained results are discussed and compared 
to other calculations in Sec.\ \ref{OV}. An alignment scheme with respect to the average 
electrostatic potential is introduced and found to bring all the calculated results in  good 
agreement with each other. The significance of this alignment of bulk band structures is discussed in more detail
in Sec.\ \ref{Alignment}. To understand our findings about defect charge transition levels, 
fundamental differences between localized and extended states in approximate 
DFT formulations are discussed in Sec.\ \ref{States}. In Sec.\ \ref{Edge},
two different theories reproducing the experimental band gap but differing in the positions 
of the bulk band edges with respect to the vacuum level are taken under consideration to 
complete our rationale.  We summarize our work and draw conclusions in Sec.\ \ref{Conclusions}.


\section{Computational methods \label{Comp}}

In the present calculations, the electronic structure was treated using two different functionals. First, we employed the
GGA functional proposed by Perdew, Burke, and Ernherhof (PBE). \cite{Perdew_PRL_1996} 
For comparison with previous calculations in the literature, we obtained for bulk ZnO a 
band gap of 0.83 eV, to be compared with the experimental value of 3.44 eV. 
To obtain an improved band gap, we used a hybrid density functional
\cite{Becke_JCP_1993} defined by a single parameter $a$ corresponding to 
the fraction of nonlocal Fock exchange admixed to the GGA exchange:
\begin{equation}
E_{\text{x}}^{\text{hybrid}}=
aE_{\text{x}}^{\text{Fock}}+(1-a)E_{\text{x}}^{\text{GGA}}.
\label{hb}
\end{equation}
A hybrid functional with $a=0.25$ and with the PBE for the GGA part \cite{Perdew_JCP_1996}
is referred to as PBE0, PBEh, or PBE1PBE. For ZnO, we obtained a band gap
of 2.82 eV using this functional. The experimental band gap is reproduced with 
$a=0.32$. In the following, we refer to this functional as to PBEh-32. 
While this adjustment of $a$ is empirical, it can be justified to a certain extent.\cite{Alkauskas_PRL_2008b,Shimazaki_JCP_2009,Alkauskas_pssb_2011,Marques_2010}
It can be shown that the optimal value of $a_\text{opt}$, i.e.\ the one which reproduces the experimental band gap, is
approximately given by $a_\text{opt} \sim 1/\varepsilon_{\infty}$. Here, $\varepsilon_{\infty}$ is the electronic part of the static
dielectric constant. For a large number of materials this relationship is approximately fulfilled.\cite{Alkauskas_pssb_2011,Marques_2010}
The adjustment of $a$ can also be justified in some cases by comparison with more accurate $GW$ calculations.\cite{Alkauskas_PRL_2008b}

The main quantity that needs to be calculated is the formation energy of the oxygen vacancy in a charge state $q$, which is
given as:\cite{VanDeWalle_JAP_2004}
\begin{equation}
E_{\text{f}}^q = E_{\text{tot}}^q - E_{\text{tot,bulk}} + \mu_\text{O} + q(\varepsilon_{\text{V}}+\varepsilon_{\text{F}}).
\label{fe}
\end{equation}
Here $E_{\text{tot}}^q$ is the total energy of the defect system containing a single O vacancy in the supercell, 
$E_{\text{tot,bulk}}$ is the total energy of the host material without any defect, $\mu_\text{O}$ is the atomic chemical 
potential of oxygen, and $\varepsilon_{\text{F}}$ is the electron chemical potential. The latter is referred to the VBM
$\varepsilon_{\text{V}}$. Except for semiconductors with degenerate doping,
$\varepsilon_{\text{F}}$ varies between zero and the band gap of the material $E_{\text{g}}$.

The atomic chemical potentials $\mu_\text{O}$ and $\mu_{\text{Zn}}$ are bound 
by the condition that ZnO is in thermal equilibrium with the reservoir of O and Zn atoms,
i.e.\ $\mu_{\text{Zn}}+\mu_{\text{O}} = \mu_{\text{ZnO}}$. Oxygen-rich conditions are defined by the
onset of spontaneous formation of O$_2$ molecules, i.e.\ by $\mu_{\text{O}} = \tfrac{1}{2}E_{\text{tot}}^{\text{O}_2}$.
Oxygen-poor (Zn-rich) conditions are correspondingly defined by the onset of spontaneous formation of bulk Zn crystallites,
i.e.\ via $\mu_{\text{Zn}}=\mu_{\text{Zn,bulk}}$. The formation of oxygen vacancies in ZnO is hindered in O-rich 
conditions, and facilitated in O-poor conditions. 
The calculation of the O chemical potential in O-poor conditions poses some difficulties when hybrid density 
functionals are used, because this involves the calculation of the total energy of bulk Zn. In Hartree-Fock
theory the description of metals leads to divergences, and the same problem is also found with hybrid functionals. 
To overcome this problem, we assume that the cohesive energy of bulk Zn, which is well described in the GGA, 
does not change significantly in the hybrid functional calculation.\cite{Carvalho_PRB_2009}
Alternatively, one could define the O chemical potential in O-poor conditions by assuming that the 
separation between the O-rich and O-poor chemical potentials in GGA is preserved in the hybrid 
functional calculation; 
this condition corresponds to assuming equal formation energies for ZnO 
in GGA and in the hybrid functional scheme.
These two ways of determining the O chemical potential in O-poor conditions lead to formation energies differing 
by about 0.4 eV.

Charge transition levels correspond to the specific value of the electron chemical potential for which two
charge states have equal formation energies. The (+2/0) charge transition level is thus given by:
\begin{equation}
\varepsilon(+2/0) = \frac{E^{0}_\text{tot} - E^{+2}_\text{tot}}{2} - \varepsilon_{\text{V}}.
\label{ctl1}
\end{equation}
Charge transition levels do not depend on atomic chemical potentials.

The calculations were performed within a plane-wave pseudopotential formulation. Soft
norm-conserving pseudopotentials \cite{Troullier_PRB_1991} were generated at the PBE level
and used in all subsequent calculations. The plane-wave kinetic energy cutoff, determined
by the much harder O pseudopotential, was set to 80 Ry. 
The calculations in the present paper were performed with the 
code {\sc cpmd}.\cite{CPMD,Hutter_CPC_2005,Todorova_JPCB_2006,Broqvist_PRB_2009}
We explicitly treated the singularity in the nonlocal exchange
potential.\cite{Broqvist_PRB_2009}

We used the experimental lattice 
parameters for bulk ZnO, since these were found to be very close to theoretical 
lattice parameters obtained with hybrid functionals.\cite{Oba_PRB_2008}
We also used experimental lattice constants in our GGA calculations, finding
results which did not differ in any significant way from GGA calculations
performed with theoretical lattice parameters.\cite{Janotti_PRB_2007} 
Upon defect formation, geometry relaxations were performed with both the GGA 
and the PBEh-32 functionals.  
The defect structures achieved in the two cases were found to be very similar: in PBEh-32, for example, 
PBE-optimized defect structures are only 0.08 eV higher in energy than those optimized consistently
at the PBEh-32 level. Hence, geometry optimization at the PBEh-32 level has no effect on the position 
of the (+2/0) charge transition level [Eq.\ (\ref{ctl1})].

\begin{figure}
\includegraphics[width=7cm]{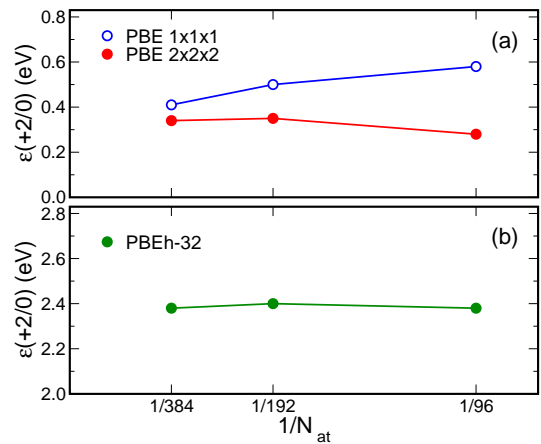}
\caption{(Color online) Charge transition level $\varepsilon(+2/0)$ vs
inverse number of atoms contained in the supercell $N_\textrm{at}$, 
(a) for the PBE calculation ($1\times1\times1$ and $2\times2\times2$ 
$k$-point meshes) and (b) for the PBEh-32 calculation ($1\times1\times1$ 
mesh). $\varepsilon(+2/0)$ is referred to the respective VBM.
\label{conv} }
\end{figure}

For the defect structures we used the supercell approach. This gives rise 
to finite-size effects which need to be accounted for. 
First, as suggested by Van de Walle and Neugebauer, \cite{VanDeWalle_JAP_2004}
the total energies of charged defects were corrected by $q\Delta V$, $\Delta V$ 
being the shift needed to align the local potential of the \emph{neutral} system 
far from the defect to that of a separate unperturbed bulk calculation, which was used to
determine $\varepsilon_{\text{V}}$. This term was found to be quite small for 
the supercells employed in our calculations. Second, the total energies of charged 
defect states are subject to spurious electrostatic contributions associated 
to the periodic boundary conditions and to the compensating background charge in our 
supercell calculations.  
To evaluate these effects, we used an extrapolation scheme based on supercell 
calculations of increasing size, containing 96, 192, and 384 atoms, as shown in Fig.\ \ref{conv}. 
When using the PBE functional, the convergence of formation energies 
and charge transition levels is accelerated when using the $2\times2\times2$ Monkhorst-Pack mesh instead 
of a sampling at the sole $\Gamma$ point [Fig.\ \ref{conv}(a)]. Hence, finite-size corrections 
are sizeable for the PBE calculation and a careful extrapolation
of the results is needed, as previously shown by Oba \emph{et al.} \cite{Oba_PRB_2008} 
At variance, a denser $k$-point mesh turned out to be unnecessary for a calculation 
with the hybrid functional PBEh-32 [Fig.\ \ref{conv}(b)].  Indeed, in the latter case, 
the bulk band gap is substantially larger and the dispersion of the defect state is already 
negligible for the smallest supercells considered. This behavior is in line with observations 
in a previous study on defects in ZnO.\cite{Carvalho_PRB_2009}
A notable difference between finite size effects in PBE and PBEh-32 calculations
suggests that unphysical defect-defect interactions mediated by bulk bands 
could be operative in the former case.\cite{Lany_PRB_2008}
For the largest supercell considered here, we obtain a conservative
estimate of 0.20 eV for the residual finite-size error by considering 
the monopole correction proposed by Makov and Payne.\cite{Makov_PRB_1995}




\section{Oxygen vacancy in Z\lowercase{n}O} \label{OV}

For the neutral oxygen vacancy, we obtained, at the PBE level, formation energies of 3.17 eV in O-rich 
conditions and of 0.50 eV in O-poor conditions. In the PBEh-32 calculation,
the corresponding value is 3.57 eV in O-rich conditions. 
In O-poor conditions, we found 0.50 and 0.90 eV depending on 
whether the cohesive energy of Zn or the formation energy of ZnO is taken from the GGA, 
respectively. Our values agree well with the value of 0.8 eV found in Ref.\ \onlinecite{Lany_PRL_2007}
and that of 0.9-1.0 eV in Ref.\ \onlinecite{Oba_PRB_2008}. Thus, our results confirm 
that the formation energy of the O vacancy in O-poor conditions 
is small enough to lead to a noticeable concentration of these defects. 

At variance with these results, Janotti and Van de Walle reported much 
higher formation energies for the neutral V$_{\text{O}}$.\cite{Janotti_PRB_2007}
They used an extrapolation procedure based on LDA+$U_{d}$ and an additional assumption about the 
behavior of the formation energy of the \emph{charged} vacancy upon the band-gap correction.
While the former extrapolation has been criticized due to the unphysical values 
to which the $U_{d}$ parameter extrapolates to,\cite{Lany_PRB_2008} we argue here that it is the 
latter assumption that is inconsistent with the hybrid functional calculations. Indeed,
the extrapolation procedure adopted in Ref.\ \onlinecite{Janotti_PRB_2007} 
leads to charge transition levels that agree well with those obtained with hybrid functionals.

\begin{figure}
\includegraphics[width=5cm]{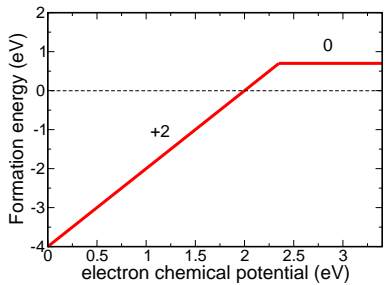}
\caption{(Color online) Formation energy of oxygen vacancy in ZnO vs electron chemical potential,
as obtained with the PBEh-32 functional. O-poor conditions are assumed. 
\label{format} }
\end{figure}

The dependence of the formation energy on the electron chemical potential is shown Fig.\ \ref{format} for 
oxygen-poor conditions. For simplicity, the oxygen chemical potential was set to
the average value derived from the two definition schemes described above. 
The (+2/0) charge transition level occurs at $\varepsilon_{\text{V}}+2.38$ eV.
This result agrees well with other calculations based on hybrid functionals.
Oba \emph{et al}.\ found the (+2/0) charge transition level at $\varepsilon_{\text{V}}+2.23$ eV,\cite{Oba_PRB_2008}
using the Heyd-Scuseria-Ernzerhof (HSE) hybrid functional based on screened exchange\cite{Heyd_JCP_2003} 
in which the fraction of non-local exchange was set to 0.375 (HSE-37.5). 
Using the same functional but with $a$ set to 0.40 (HSE-40), Clark \emph{et al}.\  
obtained the transition level at $\varepsilon_{\text{V}}+2.34$ eV.\cite{Clark_PRB_2010} 
Thus, it appears that when $a$ in either PBEh or HSE functionals is tuned to reproduce the experimental band gap, 
one consistently obtains the (+2/0) charge transition level at 2.23$-$2.38 eV from the VBM. 
The occurrence of such an agreement has recently been rationalized in general terms.\cite{Komsa_PRB_2010}
Janotti and Van de Walle,\cite{Janotti_PRB_2007} who adopted an extrapolation method based on LDA+$U_{d}$, 
found this charge transition level at 2.17 eV, in a fair agreement with the hybrid functional calculations.

As already noted in the literature, \cite{Lany_PRB_2008,Paudel_PRB_2008,Clark_PRB_2010}
the charge transition level at $\varepsilon_{\text{V}}+2.2$-$2.4$ eV is in stark 
disagreement with calculations based on other methods for correcting the band gap.  
For example, adopting a LDA$+U_{d}$ scheme,\cite{comment_Lany} Lany and Zunger 
obtained the charge transition level at $\varepsilon_{\text{V}}+1.3$ eV.\cite{Lany_PRL_2007}  
In the LDA$+U_{d}$ method, the Hubbard $U_{d}$ term acts on the Zn 3$d$ states and 
the band-gap problem is not fully corrected. When one tunes the $U_{d}$ parameter so that 
the position of Zn 3$d$ states are correctly positioned with respect to the VBM, one 
obtains a band gap of 1.5 eV, considerably smaller than the experimental one. 
The remaining band-gap error was corrected by an upward shift of the CBM.\cite{Lany_PRL_2007}

In another study, Paudel and Lambrecht adopted a LDA$+U_{s/d}$ scheme, in which the Hubbard $U$ 
term was applied to both Zn 3$d$ and Zn 4$s$ states.\cite{Paudel_PRB_2008} While
this scheme brings the theoretical band gap in agreement with experiment, 
the (+2/0) charge transition level is found at $\varepsilon_{\text{V}}+0.8$ eV. 
Some of the results obtained in Ref.\ \onlinecite{Paudel_PRB_2008} have recently
been reviewed and improved by Boonchun and Lambrecht.\cite{Boonchun_pssb_2011}
We here mainly elaborate on the original results, but the conclusions that we draw
are independent of this choice. Using a similar method as that of Paudel and Lambrecht, 
Lany and Zunger have obtained a level at $\varepsilon_{\text{V}}+0.6$ eV. \cite{Lany_PRB_2008}

\begin{figure}
\includegraphics[width=8.5cm]{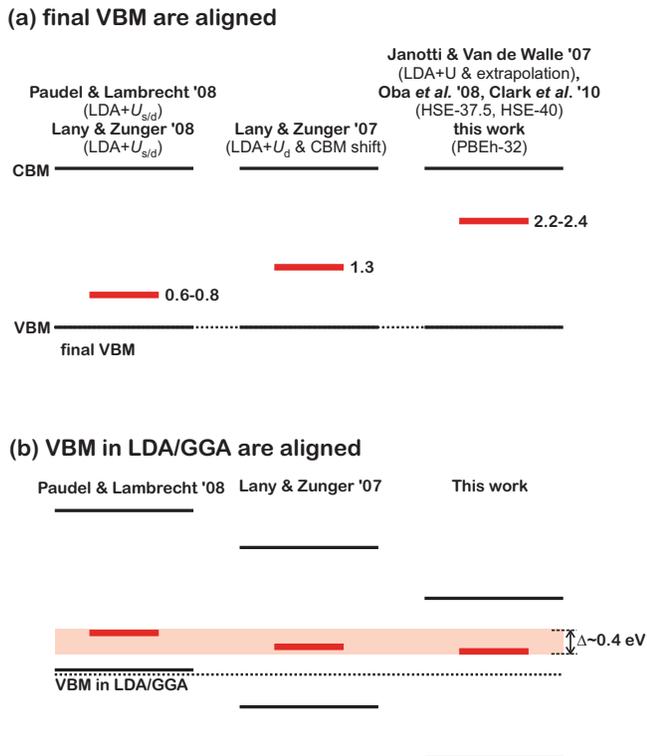}
\caption{(Color online) Calculated positions of the (+2/0) charge transition level of the oxygen vacancy in ZnO through
different band-gap correction methods: (a) the various calculations are aligned via the VBM after the band gap correction are applied;
(b) all the calculations are aligned through the VBM prior to shifts of the band edges required for the band gap correction. 
The illustrated results are taken from:
``Paudel \& Lambrecht '08" - Ref.~\onlinecite{Paudel_PRB_2008},
``Lany \& Zunger '08" - Ref.~\onlinecite{Lany_PRB_2008}, 
``Lany \& Zunger '07'' - Ref.\ \onlinecite{Lany_PRL_2007},
``Janotti \& Van de Walle '07" - Ref.~\onlinecite{Janotti_PRB_2007},
``Oba \emph{et al}.\ '08" - Ref.~\onlinecite{Oba_PRB_2008}, 
``Clark \emph{et al}.\ '10'' - Ref.\ \onlinecite{Clark_PRB_2010}.
The theoretical method is indicated in parentheses.
}
\label{ov}
\end{figure}

The charge transition levels obtained with different methods are compared in Fig.\ \ref{ov}(a).
We note that the observed differences do not stem from different electron
densities of the defect state, as the oxygen vacancy is characterized by a fully symmetric 
state of $a_1$ symmetry which is well described in all schemes. 
The origin of this apparent disagreement between various methods has lately been
discussed to some extent.\cite{Lany_PRB_2008} However, it remains unclear whether the observed  
differences originate from failures of some specific methods or whether they point to a more 
fundamental problem common to all approximate electronic structure methods. 

A clue to the understanding why different methods seemingly differ so much is provided
by the realization that the band edges of bulk ZnO calculation undergo drastically different 
shifts when going from LDA/GGA calculations \cite{LDA-GGA} to band-gap corrected schemes.  
Such shifts between two different electronic structure calculations are properly defined through 
the alignment of the average electrostatic potential. For example, the LDA$+U_{d}$ method 
of Ref.\ \onlinecite{Lany_PRL_2007} yields a shift in the VBM, $\Delta \varepsilon_{\text{V}}=-0.7$ eV. 
The LDA$+U_{s/d}$ method of Ref.\ \onlinecite{Paudel_PRB_2008} gives a shift of $+0.1$ eV, while 
our calculations yield $-1.8$ eV.

In Fig.\ \ref{ov}(b) we show the comparison of the (2+/0) charge transition level obtained
with various methods, when the VBMs in the LDA/GGA calculations are aligned. 
This is equivalent to aligning the electrostatic potential of all calculations
(see Sec.\ \ref{Alignment}). With this alignment, the various methods
yield charge transition levels differing by at most 0.4 eV. This is to be contrasted
to the variation of up to 1.8 eV achieved when the electronic structures are aligned via 
their respective VBM [Fig.\ \ref{ov}(a)]. Thus, these theoretical calculations 
do not in fact differ as much as has been previously claimed.
Our conclusion is that, when a suitably defined common reference level is adopted, 
the charge transition levels are more accurately described than the bulk band 
edges.\cite{Alkauskas_PRL_2008a} In Secs.\ \ref{States} and \ref{Edge} below, 
we give a detailed explanation of this behavior and address its general 
consequences for theoretical studies of defects.


\section{Alignment of bulk band structures\label{Alignment}}

The previous discussion relied on the assumption that the bulk band structures 
of two theoretical calculations can be aligned with respect to each other, 
as done in Fig.\ \ref{ov}(b). This alignment allows one to determine 
the shifts in the valence band $\Delta \varepsilon_{\text{V}}$ and 
in the conduction band $\Delta \varepsilon_{\text{C}}$ for a given 
theoretical scheme with respect to another one. In this section, 
we discuss the meaning of such an alignment.\cite{Alkauskas_PRL_2008a,Alkauskas_pssb_2011}

The alignment between the electronic structures of the same bulk material  
within different theoretical schemes could in principle be 
achieved through the identification of a common reference potential. 
For instance, the vacuum level could serve this purpose, requiring 
the explicit consideration of the surface between the considered material
and vacuum within both theoretical schemes. 
Since the surface dipole depends on the specific crystal surface which is considered,
the same orientation has to be chosen for both theoretical schemes.
In this way, properly defined 
bulk levels in the two schemes, such as $\varepsilon_{\text{V}}$ and 
$\varepsilon_{\text{C}}$, can be positioned with respect to the vacuum level
and thus aligned. By constructon, the alignment achieved in this way is {\it not}
an intrinsic bulk property of the two theoretical schemes. Indeed,
differences between the surface dipoles in the two surface calculations
directly affect the alignment.

While such a procedure can always be carried out, we note that the 
alignment between different electronic structures for the same bulk material is 
a meaningful concept only as long as their associated electron densities are
identical (or very close).  Indeed, different electron densities at surfaces 
of the material could yield different surface dipoles and thus the achieved 
alignment would depend on the particular surface adopted and give rise to ambiguity.
Moreover, different surface dipoles could result from different electron 
densities in the bulk, for instance because of different theoretical equilibrium 
lattice parameters. In such a case, the alignment with respect to the vacuum level 
would again be surface dependent. When comparing electronic structures of bulk 
materials as achieved within different theoretical schemes, we will thus 
additionally assume that their electron densities do not differ essentially.
In practical calculations involving semilocal and hybrid density functionals,
this condition is close to being satisfied. Indeed, surface and interface dipoles 
in a variety of cases were found to differ by at most a few tenths of an eV.%
\cite{Alkauskas_PRL_2008b,Broqvist_APL_2008,Broqvist_APL_2009,Lyons_PRB_2009,Komsa_PRB_2010}

Under the assumption of yielding close electron densities, two different 
theoretical schemes can be expected to give similar surface dipoles. This implies 
that an alignment to the vacuum level is equivalent to an alignment 
to the average electrostatic potentials within the bulk of the materials.\cite{Alkauskas_pssb_2011}
This consequence is particularly convenient and allows us to compare different
bulk calculations without the necessity of performing surface calculations.\cite{Alkauskas_PRL_2008a}
Note, however, that it is implicitly understood that alignment shifts 
resulting from slight differences in the electron density 
are negligible when compared to the shifts undergone by the band edges.

To produce Fig.\ \ref{ov}(b), we relied on shifts $\Delta \varepsilon_{\text{V}}$ and $\Delta \varepsilon_{\text{C}}$
calculated in the respective papers. Indeed, the position of the VBM and the CBM in the more advanced theory 
were generally given with respect to the (semi-)local density functional calculation (LDA or GGA) for an 
alignment with respect to the average electrostatic potential. 
For instance, the LDA$+U_{s/d}$ and LDA band structures obtained in Ref.\ \onlinecite{Paudel_PRB_2008}, 
corresponding to the left column in Fig.\ \ref{ov}(b), were aligned through the average electrostatic potential
in the bulk. In Refs.\ \onlinecite{Lany_PRL_2007, Lany_PRB_2008}, corresponding to the results in the middle column in Fig.\ \ref{ov}(b),
the authors determined the shifts of the bulk bands in the LDA$+U_{d}$ with respect to the LDA by referring 
the energies to O 2$s$ states which do not directly couple to the $d$ states on which the Hubbard correction was applied.
This is again equivalent to the alignment to the average electrostatic potential in the bulk.
In our own calculations, presented in the right column in Fig.\ \ref{ov}(b), we aligned the two band structures
through the average electrostatic potential in the bulk.
Unfortunately, the reported data did not allow us to establish the relative alignment 
for all the studies referred to in Fig.\ \ref{ov}(a). However, we can assume that similar
theories yield close $\Delta \varepsilon_{\text{V}}$ and $\Delta \varepsilon_{\text{C}}$. 
For instance, the LDA$+U_{s/d}$ calculations of Lany and Zunger \cite{Lany_PRB_2008} are expected to
yield similar shifts as those found by Paudel and Lambrecht \cite{Paudel_PRB_2008} [Fig.\ \ref{ov}(a)]. 
As far as the screened hybrid functionals are concerned [Fig.\ \ref{ov}(a)], a recent study has shown 
that these functionals yield very similar shifts as the unscreened functionals used in our calculations, as 
long as the fraction of nonlocal exchange is tuned to reproduce the experimental band gap.\cite{Komsa_PRB_2010} 
Hence, although the results in Fig.\ \ref{ov}(b) are restricted to those studies which 
explicitly give the shifts in the band edges, the present considerations are expected to 
carry a much broader validity and to equally hold for all other calculations reported in Fig.\ \ref{ov}(a).


\section{Localized and delocalized states in approximate density functional schemes \label{States}}

We showed above that different theoretical models give quite consistent results concerning
the description of the (+2/0) charge transition level of the O vacancy in ZnO provided they 
are aligned through the average electrostatic potential, taken as a common reference level. 
To understand why this happens, we first discuss fundamental differences between 
localized (atomic-like) and extended (bulk-like) states in approximate density functional schemes.

For (approximate) density functionals Janak's theorem \cite{Janak_PRB_1978} applies:
\begin{equation}
\frac{\partial E_{\text{tot}}}{\partial f_i} = \varepsilon_i(f_i),
\label{Janak1}
\end{equation}
i.e.\ the derivative of the total energy with respect to the 
change of occupation number $f_i$ of the highest-occupied state
$i$ is equal to the single-particle eigenvalue of this state 
$\varepsilon_i$, when the latter is referred to the average 
local potential.\cite{Ihm_JPC_1979}

The integral form of Janak's theorem is
\begin{equation}
E^{N}_{\text{tot}}-E^{N-1}_{\text{tot}} = \int_0^1 \varepsilon_N (f) df,
\label{Janak2}
\end{equation}
where $E^{N}_{\text{tot}}$ is the total energy of the system 
with $N$ electrons. In the above expressions, we suppressed the spin variable. 
While in the original derivation of Janak's theorem the functionals were implicitly assumed 
to be continuous, Eq.\ (\ref{Janak1}) equally applies to functionals which possess a 
discontinuity \cite{Perdew_PRL_1982,Perdew_PRL_1983} at integer number of electrons. In this case one has 
to distinguish between left and right derivatives and the corresponding single particle eigenvalues. 
The integral form of Janak's theorem, Eq.\ (\ref{Janak2}), applies to discontinuous functionals without
modifications.


In the case of localized states, such as, e.g., in molecules and atoms, the single particle eigenvalue
in approximate density functional schemes depends sensitively on the fractional occupation. 
Accordingly, total energy differences pertaining to the change of number of electrons are given 
by Eq.\ (\ref{Janak2}). In particular, the ionization potential (IP) of a 
system 
is given by
\begin{equation}
\text{IP} = E^{N-1}_{\text{tot}}-E^{N}_{\text{tot}} = -\int_0^1 \varepsilon_N(f) df,
\label{Janak_IP}
\end{equation}
where $\varepsilon_N$ is the highest occupied orbital of the $N$-electron system. 
Similarly, the electron affinity (EA) can be expressed as
\begin{equation}
\text{EA} = E^{N}_{\text{tot}} - E^{N+1}_{\text{tot}} = -\int_0^1\varepsilon_{N+1}(f) df,
\label{Janak_EA}
\end{equation}
where $\varepsilon_{N+1}$ is the lowest unoccupied state of the $N$-electron system.

It has been known for some time that total energy differences pertaining to the change of charge state of a localized
state are quite accurately described in approximate density functional schemes, both in semilocal and hybrid ones. 
For example, Curtiss \emph{et al}.\ calculated IPs and EAs for a large set of molecules 
using GGA and hybrid functionals.\cite{Curtiss_JCP_1997} They calculated these quantities  
via total energy differences ($\Delta$SCF method), yielding an average deviation with respect 
to experiment lower than $0.2$-$0.3$ eV for both GGA (BLYP) and hybrid (B3LYP) functionals.
This accuracy is achieved despite the fact that the single-particle eigenvalues of the highest-occupied
molecular orbital (HOMO) $\varepsilon_{\text{HOMO}}$ and of the lowest-unoccupied molecular orbital (LUMO)
$\varepsilon_{\text{LUMO}}$ are substantially different in the GGA and in hybrid functional schemes.
A similar agreement with experiment also holds for screened hybrid functionals.\cite{Heyd_JCP_2004} 
However, plain LDA yields slightly larger errors, of the order of $0.5$-$0.6$ eV 
for the same quantities.\cite{Curtiss_JCP_1997}

\begin{figure}
\includegraphics[width=8.5cm]{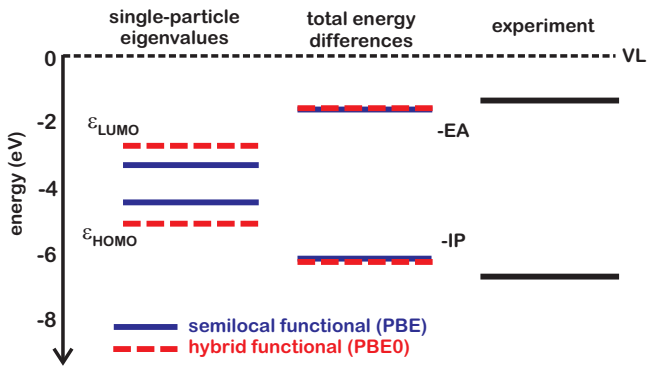}
\caption{(Color online) Frontier-orbital diagram of the pentacene molecule. Left panel: 
HOMO and LUMO single-particle eigenvalues as obtained with the semilocal PBE functional (dashed, blue) and
with the hybrid PBE0 functional (solid, red). Middle panel: Ionization potentials and electron 
affinities calculated with the same functionals. Right panel: Experimental values for the electron
affinity and the ionization potential.
}
\label{molecules}
\end{figure}

We illustrate this property in the case of the pentacene (C$_{22}$H$_{12}$) molecule 
in Fig.\ \ref{molecules}.\cite{note_pentacene} Pentacene is a convenient example because, 
unlike several smaller acenes, it possesses a positive electron affinity.
The single-particle HOMO and LUMO levels, calculated with the semilocal PBE functional 
(left, solid lines), do not agree well with the negative of the experimental IP and EA (right).  
In particular, the single-particle gap 
$E_{\text{g}}^{\text{KS}}=\varepsilon_{\text{LUMO}}-\varepsilon_{\text{HOMO}}$ of 1.12 eV
is severely underestimated with respect to the experimental gap $E_{\text{g}}=\text{IP}-\text{EA}$ of 5.29 eV.
The use of the hybrid PBE0 (i.e.\ PBEh-25) functional (left, dashed lines) gives some improvement, but
the calculated single-particle HOMO-LUMO gap of 2.34 eV remains much smaller than the experimental one.
At variance, when calculated via total energy differences, the IPs and EAs in both PBE and PBE0 
are much closer to their corresponding experimental values, 6.64 eV\cite{Clark_HCA_1972} 
and 1.35 eV,\cite{Crocker_JACS_1993} respectively. The two theoretical values (Fig.\ \ref{molecules}) 
differ by less than 0.10 eV, with the hybrid functional calculation in slightly better agreement with the experimental results. 
The residual differences between calculated and measured values ($\sim$0.45 eV for the IP and $\sim$0.25 eV for the EA) 
can be accounted for by the quite large electron correlation effects in the pentacene molecule.\cite{Deleuze_JCP_2003} 
In any case, the present result shows that these theoretical schemes yield total energy differences
in good agreement with experiment and with each other, while the single-particle levels in the two schemes
are very different. This is consistent with the general trend found by Curtiss \emph{et al.} 
for a large set of smaller molecules.\cite{Curtiss_JCP_1997}

Thus, we conclude that total energy differences pertaining to the change of charge state of 
localized states are accurately described with approximate density functionals. Approximating the 
integrals appearing in Eqs.\ (\ref{Janak_IP}) and (\ref{Janak_EA}) through the trapezoidal 
rule, we arrive at the following expressions for the IP and the EA:
\begin{equation}
\text{IP}=E^{N-1}_{\text{tot}}-E^{N}_{\text{tot}} \approx -\varepsilon_N\left(\tfrac{1}{2}\right)
\label{Slater_IP}
\end{equation}
and
\begin{equation}
\text{EA}=E^{N}_{\text{tot}}-E^{N+1}_{\text{tot}} \approx -\varepsilon_{N+1}\left(\tfrac{1}{2}\right).
\label{Slater_EA}
\end{equation}
Here, $\varepsilon_{N}=\varepsilon_{\text{HOMO}}$ and $\varepsilon_{N+1}=\varepsilon_{\text{LUMO}}$.
Electronic states at half-filling correspond to Slater-transition states.\cite{Slater}
Since Eqs.\ (\ref{Slater_IP}) and (\ref{Slater_EA}) apply equally well to various semilocal and hybrid functionals,
the generally good agreement with experiment implies that the respective eigenvalues $\varepsilon(f)$ defined as a 
function of filling  all approximately cross at half-filling. This has indeed already been observed.\cite{Vydrov_JCP_2007}

The reason for this good performance of approximate density functionals should be
ascribed to the fact that such functionals fulfill several exact constraints
of the many-body fermionic system.\cite{Perdew_LNP_2003} In particular,
the most relevant in this context is the generalized sum-rule of the exchange-correlation hole.
This rule holds for systems with an \emph{integer} number 
of electrons, i.e.\ for closed systems in which no exchange of electrons with the environment occurs.\cite{Perdew_PRB_1997}
This condition is enforced for most approximate functionals, including the LDA and various GGAs. Furthermore,
since this constraint is naturally fulfilled in the Hartree-Fock theory, it also holds for any hybrid functional 
with an exchange energy of the type given in Eq.\ (\ref{hb}).

\begin{figure}
\includegraphics[width=8.5cm]{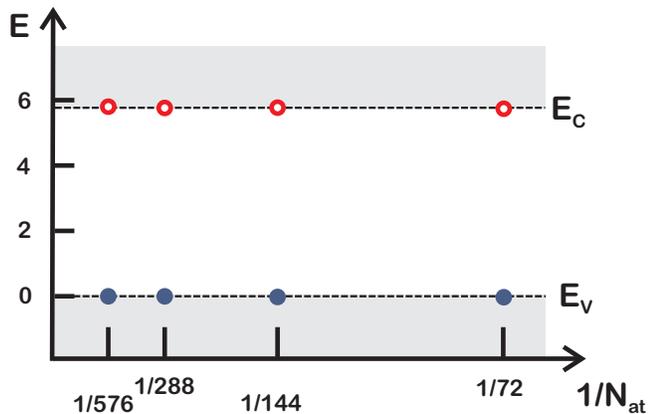}
\caption{(Color online) Band edges of crystalline SiO$_2$ ($\alpha$-quartz) calculated via total energy differences
as a function of $1/N_{\text{at}}$, where $N_{\text{at}}$ is the total number of atoms
in the supercell. Calculations have been performed with the semilocal PBE functional.}
\label{extended}
\end{figure}

The situation is very different in the case of infinitely extended bulk-like states. Indeed the band-gap problem
pertaining to (generalized) Kohn-Sham eigenvalues {\it cannot} be overcome by considering total-energy differences. 
When a fraction $f$ of an electron or even a full electron is added to or removed from an extended state, the
total electron density changes negligibly. Thus, the local potential, both the Hartree and the approximate exchange-correlation 
potential 
remain unaffected. 
As a result, the single-particle eigenvalues do not depend on the filling $f$ of this state. Using the
integral form of Janak's theorem given in Eq.\ (\ref{Janak2}), we get for the valence band maximum:
\begin{equation}
\varepsilon_{\text{V}}=E^{N}_{\text{tot}}-E^{N-1}_{\text{tot}},
\label{TED_EV}
\end{equation}
and for the conduction band minimum:
\begin{equation}
\varepsilon_{\text{C}}=E^{N+1}_{\text{tot}}-E^{N}_{\text{tot}}.
\label{TED_EC}
\end{equation}
To illustrate this property, we show in Fig.\ \ref{extended} the VBM and CBM of $\alpha$-quartz
calculated via total energy differences as a function of the supercell size. The considered cells contain
72, 144, 288, and 576 atoms, and their Brillouin zones are sampled at the sole $\Gamma$ point.
The semilocal PBE functional was used. In the case of $\alpha$-quartz, total energy differences
are very close to single particle eigenvalues already for the smallest cells.
For the 72-atom cell, the difference is 0.015 eV for the VBM and 0.035 eV for the CBM, while for the 576 cell these
are 0.003 eV and 0.004 eV, respectively. 
The particular case of hybrid functionals has been addressed in detail in Ref.\ \onlinecite{Broqvist_PRB_2009}.
Hence, unlike for the localized states in Fig.\ \ref{molecules}, the consideration of total-energy differences in the
case of extended states is not useful to improve the comparison with experiment and the same limitations pertaining 
to the single-particle eigenvalues (band-gap problem) are encountered. \cite{Foster_PRB_2001,Pruneda_PRB_2005}  
A similar comparison involving extended states of GaAs and localized states of the F atom can be found in 
Ref.\ \onlinecite{Lany_PRB_2008}.

The above discussion highlights an important difference between localized and extended states as 
described within approximate density functional schemes. While the band-gap problem associated to 
single particle eigenvalues can be circumvented by considering total-energy differences for localized 
states, such a solution does not apply to extended states for which the band-gap problem 
remains a fundamental obstacle. 
Recently, a clear explanation has been put forward for justifying this
different behavior.\cite{Mori-Sanchez_PRL_2008, Cohen_PRB_2008}
The inaccurate total energies for large systems with \emph{integer} number 
of electrons stems from the failure of approximate density functionals in 
describing small systems with \emph{fractional} 
charges.\cite{Mori-Sanchez_PRL_2008, Cohen_PRB_2008} 
Indeed, approximate functionals generally do not reproduce the property of the exact 
density functional by which the total energy depends linearly on the number of electrons. 
There is at present an on-going effort to achieve improved descriptions on the
basis of these ideas.\cite{Lany_PRB_2009,Lany_PRB_2010b,Dabo_PRB_2010}
The degree of localization required for achieving an accurate description 
with current density functionals is still to a large extent an 
open question. We refer the reader to the interesting debate 
on this issue in Refs.\ \onlinecite{Ogut_PRL_1998}.


\section{``The band-edge problem''\label{Edge}}

Having stressed the different properties of localized and extended states with respect to a change in electron occupation,
we return in this section to the discussion of charge transition levels. For the sake of simplicity, let us
consider the $(+/0)$ transition of a point defect characterized by an atomically localized wave function. 
Using Eq.\ (\ref{TED_EV}), we write the charge transition level $\varepsilon(+/0)$ as the difference 
between two terms, each of them corresponding to a total-energy difference:
\begin{eqnarray}
\varepsilon(+/0)& = & E_{\text{tot}}^{0} - E_{\text{tot}}^{+} - \varepsilon_{\text{V}} \nonumber \\
&=& \underbrace{\left(E_{\text{tot}}^{0} - E_{\text{tot}}^{+}\right)}_{\text{~ localized state}}
- \underbrace{\left(E_{\text{tot, bulk}}^{0} - E_{\text{tot,bulk}}^{+}\right)}_{\text{delocalized state}}.
\nonumber \\
\label{TED2}
\end{eqnarray}
The second term clearly describes the total energy difference pertaining to a delocalized bulk state,
while the first term can to a very good approximation be related to the total energy difference pertaining 
to a localized state. Formally, the first term describes the total energy of the whole manifold 
of states involving both defect and bulk states, but can be related to the localized defect state 
through the Slater transition-state approximation.\cite{Alkauskas_PRL_2008a} For atomically localized 
defect states, this is a very good approximation.\cite{Alkauskas_PB_2007} 
In view of the following discussion, it is convenient to rewrite Eq.\ (\ref{TED2}) as
\begin{eqnarray}
\varepsilon(+/0)&=& \left( E_{\text{tot}}^{0} - E_{\text{tot}}^{+} - \phi \right)- \left (\varepsilon_{\text{V}}-\phi \right) \nonumber \\
&=&\bar{\varepsilon}(+/0) - \bar{\varepsilon}_{\text{V}}, 
\label{TED3}
\end{eqnarray}
where the charge transition level $\bar{\varepsilon}(+/0)$ and the VBM $\bar{\varepsilon}_{\text{v}}$ 
are referred to the average electrostatic potential $\phi$ of the unperturbed bulk material.

\subsection{``Band-gap'' problem of defect energy levels \label{band-gap}}

Let us assume that we study the $(+/0)$ charge transition level of the same defect using 
two different theories: theory I and theory II. The first theory severely underestimates 
the band gap, while the second one gives a band gap in a much closer agreement with experiment. 
The two theories differ only by the exchange-correlation potential. According to Eq.\ (\ref{TED3}),
the corresponding charge transition levels referred to the respective valence band maxima are:
\begin{equation}
\varepsilon^{\text{I}}(+/0) = \bar{\varepsilon}^{\text{I}}(+/0) - \bar{\varepsilon}^{\text{I}}_{\text{V}},
\label{rw1}
\end{equation}
and
\begin{equation}
\varepsilon^{\text{II}}(+/0) = \bar{\varepsilon}^{\text{II}}(+/0) - \bar{\varepsilon}^{\text{II}}_{\text{V}}.
\label{rw2}
\end{equation}
We further assume that the two theories produce a sufficiently accurate representation of the electron density 
so that it is justified to align the two bulk band structures through the average electrostatic 
potential $\phi$ in the two theories, as discussed in Sec.\ \ref{Alignment}.
Under the assumption that the defect wave function $\psi_{\text{d}}$ differs very little in the two theories, 
we can express the difference between the two charge transition levels $\bar{\varepsilon}(+/0)$ making use of 
the Slater transition state:\cite{Alkauskas_PRL_2008a}
\begin{equation}
\bar{\varepsilon}^{\text{II}}(+/0)-\bar{\varepsilon}^{\text{I}}(+/0) \approx \left \langle \psi_{\text{d}} \left | \hat{V}^{\text{II}}_{\text{xc}} -
\hat{V}^{\text{I}}_{\text{xc}} \right | \psi_{\text{d}} \right \rangle,
\label{Janak_defect}
\end{equation}
where the exchange-correlation potentials are evaluated with the defect state at half occupation.
Only the difference in the exchange-correlation potentials enters the expression in Eq.\ (\ref{Janak_defect}).
Indeed, if the electron density and the single-particle wave functions are very similar in the two calculations, 
the interaction between the defect and the ionic cores, the long-range electrostatic electron-electron 
interaction, and the kinetic energy are the same in the two theories and cancel.

\begin{figure}
\includegraphics[width=8.5cm]{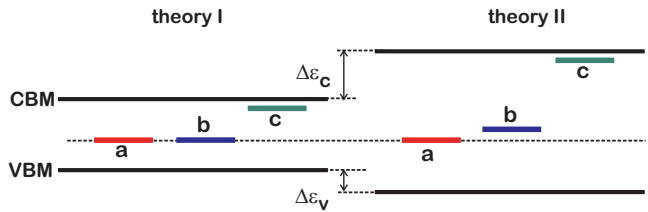}
\caption{(Color online) \label{th2} 
Schematic illustration of energy levels of various type of defect states differing 
by the extent of their wave function: (a) defect level with an atomically localized wave function,
(b) an intermediate case, and (c) an effective-mass-like defect.
The results of two electronic structure theories (theory I and theory II) giving different band gaps
are compared to illustrate the band-gap problem. The alignment is made through the 
average electrostatic potential.}
\end{figure}

To understand the behavior of defect levels, it is convenient to focus first on defects with 
extremely localized wave functions. Hence, according to Eq.\ (\ref{Janak_defect}), the 
difference $\bar{\varepsilon}^{\text{II}}(+/0)-\bar{\varepsilon}^{\text{II}}(+/0)$
can then be expressed in terms of an expectation value involving the sole localized 
defect state.\cite{Alkauskas_PRL_2008a} 
However, we know from Sec.\ \ref{States} that total energy differences
pertaining to localized states, or, equivalently, Slater transition-state eigenvalues 
of localized states, are almost the same, independent of the functional. Thus, we get:
\begin{equation}
\bar{\varepsilon}^{\text{II}}(+/0)-\bar{\varepsilon}^{\text{I}}(+/0) \approx 0.
\end{equation}
This means that charge transition levels for such defects are almost equal in the two 
theories, when the energy scales are aligned through the average electrostatic 
potential $\phi$. At variance, the charge transition levels are substantially 
different when the energy scales in the two theories are aligned through 
the respective valence band maxima, i.e.\ through $\bar{\varepsilon}_{\text{V}}$ 
in Eqs.\ (\ref{rw1}) and (\ref{rw2}), because of the different  
positions of the bulk band edges with respect to the potential $\phi$. 
This scenario pertaining to a defect with an extremely localized wave 
function is illustrated in Fig.\ \ref{th2} by the defect state $a$. 
The validity of the ideal alignment illustrated by this type of defect 
has been demonstrated for a wide class of defects encompassing 
various host materials.\cite{Alkauskas_PRL_2008a,Broqvist_APL_2008,Komsa_PRB_2010,Komsa_APL_2010,Chen_PRB_2010,Komsa_PRB_2011} 

Figure \ref{th2} also illustrates the shifts of other type of defects. In the opposite limit,
defect $c$ corresponds to an effective-mass-like defect with a spatially extended wave function.
In this case, the defect level is anchored to the bulk band to which it pertains and rigidly 
follows the band edge upon the opening of the band gap in theory II. Defect $b$ has an intermediate 
extension compared to defects $a$ and $c$, and is partially affected by the shift of the band edges. 
The relation between the departure from ideal alignment and the spatial extension of the defect wave 
functions has been documented for various defects and host materials in Ref.\ \onlinecite{Alkauskas_PRL_2008a}.
However, the detailed behavior of such defects is intrinsically system-dependent, and no universal 
considerations can be made.

In this section, we limited the discussion to defects states occurring in the band gap for 
both theories. More complex situations occur when defect states are resonant with the band states
for one of the theories.\cite{Lany_PRB_2008} However, the physical description of the defect state
is altered in such cases. The main motivation of the present work is to understand the effect 
of the band gap opening under the assumption that the defect wave function remains essentially unmodified.

\subsection{``Band-edge'' problem for defect energy levels \label{band-edge}}

In the previous section, we compared defect charge transition levels 
as obtained within two different theories giving different band gaps.
We found that the energy levels of defects states described by atomically 
localized wave functions are already well described in theories with 
a pronounced ``band-gap problem'', provided those levels are referred 
to a relevant reference level. For such defects, the problem of finding 
the defect level is essentially decoupled from that of finding the band edges.


\begin{figure}
\includegraphics[width=8.5cm]{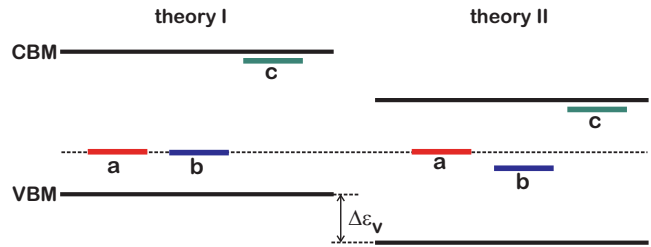}
\caption{(Color online) \label{th3}
Schematic illustration of energy levels of various type of defect states differing
by the extent of their wave function: (a) a deep defect level with an atomically localized wave function,
(b) an intermediate case, and (c) an effective-mass-like defect.
The results of two electronic structure theories (theory I and theory II) giving the same 
band gap but different band edge positions are compared to illustrate the ``band-edge problem''. 
The alignment is made through the average electrostatic potential.}
\end{figure}

Let us thus consider two different theories, theory I and theory II, yielding this time the 
same band gap (taken to coincide with the experimental one), but giving different positions 
of the VBM and the CBM with respect to the average electrostatic potential $\phi$ of the bulk. 
We assume that the two theories are sufficently accurate yielding in particular close electron 
densities, so that the energy scales of the two theories can be aligned through $\phi$, as 
discussed in Sec.\ \ref{Alignment}.
For instance, theory I could be LDA+$U$ in which the remaining band-gap underestimation 
is corrected by a rigid shift of the conduction band, while theory II could be a hybrid 
functional scheme in which the fraction of Fock exchange is tuned to reproduce the experimental 
band gap.  
For an atomically localized defect, the same argument holds as in the previous section and
the charge transition levels obtained within the two theories are expected to fall 
very close to each other, as illustrated in Fig.\ \ref{th3} for defect $a$. 
In Fig.\ \ref{th3}, a departure from the ideal alignment is seen for defects $b$ and $c$,
corresponding to defect wave functions of intermediate and effective-mass-like extensions, respectively.

Figure \ref{th3} summarizes the principal finding of the present work.
In a condensed form, the following statement can be formulated concerning the
comparison of charge transition levels of atomically localized defects. 
Despite the good description of the experimental band gap in both theories, 
the defect levels differ when referred to their respective VBM, because the 
band edges in the two theories are located differently with respect to the 
common electrostatic potential $\phi$. This occurs even when the defect wave 
function is almost identical in the two theories. This alignment property 
deteriorates with the extension of the defect wave function.
Thus, the correct description of band edges relative to the average electrostatic 
potential is a crucial prerequisite for an accurate location of charge transition 
levels within the band gap. We refer to this issue as to the ``band-edge problem'' 
for the calculation of defect levels. In other words, there is not only
a ``band gap problem'' related to the underestimation of the band gap but 
also a ``band-edge problem'' related to the position of the band edges 
with respect to the average electrostatic potential, ultimately corresponding
to an absolute alignment with respect to an external vacuum level.


As far as the determination of the (+2/0) charge transition level of the oxygen 
vacancy in ZnO is concerned, the present considerations appear confirmed [cf.\ Fig.\ \ref{ov}(b)].
This defect level behaves like the defect state $b$ in Fig.\ \ref{th3}, showing a shift
which does not depart in a significant way from the case of ideal alignment (defect state $a$).
Indeed, when referred to a common reference level, all previous calculations yield 
the (+2/0) level within 0.4 eV,\cite{Lany_PRB_2008,Janotti_APL_2005,Lany_PRL_2007,Janotti_PRB_2007,Paudel_PRB_2008,Oba_PRB_2008,Clark_PRB_2010}
which corresponds to just one 
ninth of the band gap of bulk ZnO.
Hence, contrary to previous claims, we find that all previous defect calculations 
agree quite well with each other. In fact, these calculations differ in the positions 
of the bulk band edges with respect to the average electrostatic potential.

\subsection{Which band edge shifts are the right ones?}


These considerations lead to the question about which theoretical description should be
adopted for positioning the band edges. This corresponds to determining the shift 
$\Delta \varepsilon_{\text{V}}$ of the valence band and the shift $\Delta \varepsilon_{\text{C}}$ 
of the conduction band, when taking the LDA or the GGA as a starting point. 
A direct comparison between theory and experiment is in principle possible. 
The bulk band structure can for instance be referred to the vacuum level through a surface calculation.
The VBM and the CBM determined in this way could then be compared with ionization potentials 
and electron affinities, as obtained by means of photoelectron and inverse photoelectron spectroscopy.
However, such measurements are often shrouded by very pronounced effects associated to 
charged native defects and impurities which influence the electrostatics and alter the alignment.
More practically, the validity of a given theoretical scheme can be examined addressing 
band offsets at interfaces.\cite{VanDeWalle_PRB_1987} 
Band offsets are well-defined quantities and can generally be measured through a large set
of experimental techniques.
The comparison between theoretical and experimental band offsets then
allows one to determine the overall accuracy with which such shifts are obtained within 
various theoretical schemes.\cite{Devynck_PRB_2007,Alkauskas_PRL_2008b,Shaltaf_PRL_2008,%
Broqvist_APL_2009,Komsa_PRB_2010,Gruning_PRB_2010,Wadehra_APL_2010}

In the absence of experimental data, the validity of the shifts $\Delta \varepsilon_{\text{V}}$ 
and $\Delta \varepsilon_{\text{C}}$ could also be assessed by comparing with electronic structure
calculations of higher accuracy, such as those based on many-body perturbation theory (MBPT) in
the $GW$ approximation or beyond.\cite{Alkauskas_PRL_2008b,Shaltaf_PRL_2008,Shaltaf_PRB_2009}
Indeed, such calculations not only provide improved relative positions of bulk bands, but also 
shifts of those bands with respect to theoretical schemes of lower level. 
However, recent work has shown that the shifts of the band edges with respect to
the average electrostatic potential are more difficult to converge than relative
positions of bands.\cite{Shaltaf_PRL_2008} Furthermore, such shifts are sensitive to various 
levels of approximation, such as, e.g., the use of different  models for the plasmon pole to describe
the frequency dependence of the dielectric function, the inclusion of vertex corrections $\Gamma$,
and various levels of self-consistency on $G$, $W$, $\Gamma$, and the electron wave functions.
\cite{Shaltaf_PRL_2008,Shishkin_PRL_2007,Shishkin_PRB_2007} 
To illustrate this point, we quote a recent work, \cite{Shaltaf_PRL_2008}
in which the relative shift of the valence band with respect to the overall
band gap correction, i.e.\ $\Delta \varepsilon_{\text{V}}/ \Delta E_{\text{g}}$,
was found to range from $-$0.68 to $-$0.42 in the case of SiO$_2$,
depending on the level of approximation in the $GW$ scheme.
Even for a material as simple as Si the value of 
$\Delta \varepsilon_{\text{V}}/ \Delta E_{\text{g}}$
as predicted by different $GW$ schemes ranges from $-$0.75 to +0.17.\cite{Shaltaf_PRL_2008}
Thus, clearly more work is needed to clarify these issues. A systematic study of the effects 
of different levels of approximation in MBPT on the shifts
in the band edges is thus vital for the study of defect levels. 


\section{Conclusions \label{Conclusions}}

In this work, we carried out a theoretical analysis of the (+2/0) charge transition 
level of the oxygen vacancy in ZnO. In recent years, this defect has grown into a
benchmark case to assess the quality of various advanced electronic-structure theories.
Indeed, common approximations to density functional theory, such as the LDA and the various 
GGAs, severely underestimate the band gap of bulk ZnO, and the treatment at a more advanced 
level thus becomes crucial even for drawing qualitative conclusions. However,  
different advanced theoretical methods applied hitherto yielded conflicting results 
regarding the position of the defect level in the band gap.

We here showed that apparently conflicting theoretical results are in a much better agreement with each other
when the charge transition levels are aligned with respect to the average electrostatic potential
rather than to the respective valence band maximum. We showed that the former alignment is equivalent
to the choice of a common external potential such as the vacuum level, provided the electron densities are 
sufficiently accurately described.  We have rationalized our finding by considering fundamental differences 
between the ways approximate density functionals describe localized and delocalized states. For localized states, 
the ``band-gap problem'' can generally be overcome through the consideration of total energy differences.
On the other hand, such a solution is not applicable to delocalized states, for which the 
``band-gap problem'' remains an intrinsic shortcoming.

In particular, the present study highlights a specific criterion that needs to be fulfilled in order
to properly describe charge transition level and formation energies of defects.
We clearly demonstrated that the band structure of the host material needs to be correctly 
positioned  with respect to an external potential, such as the vacuum level. When the 
electron density is accurately described, this alignment condition can equivalently 
be replaced by the alignment with respect to the average electrostatic potential in the bulk.
This condition is additional with respect to the accurate reproduction of the band gap.
Our analysis of the oxygen vacancy in ZnO shows that conflicting theoretical results arise 
for theories yielding an accurate band gap, but differing positions for the band edges. 

\section*{Ackowledgements}
We particularly thank P. Broqvist for his contribution to a 
wider research project from which the present study takes its origin.
We also acknowledge fruitful interactions with A. Carvalho, H.-P. Komsa, 
and O. A. Vydrov. Partial financial support from the Swiss National Science 
Foundation is acknowledged under Grant No.\ 200020-111747. We used 
computational resources at DIT-EPFL (BlueGene), CSEA-EPFL, and CSCS.


\begin{thebibliography}{10}

\bibitem{Stoneham}
A. M. Stoneham,
\emph{Theory of Defects in Solids: Electronic Structure of Defects in Insulators and Semiconductors}
(Oxford University Press, 1975).

\bibitem{VanDeWalle_JAP_2004}
C. G. Van de Walle and J. Neugebauer,
J. Appl. Phys. \textbf{95}, 3851 (2004).

\bibitem{Wiley}
{\it Advanced Calculations for Defects in Materials},
edited by A. Alkauskas, P. De\'ak, J. Neugebauer,
A. Pasquarello, and C. G. Van de Walle (Wiley, Weinheim, 2011).


\bibitem{Lany_PRB_2008}
S. Lany and A. Zunger,
Phys. Rev. B \textbf{78}, 235104 (2008).

\bibitem{Alkauskas_PRL_2008a}
A. Alkauskas, P. Broqvist, and A. Pasquarello,
Phys. Rev. Lett. \textbf{101}, 046405 (2008).

\bibitem{Lany_MSMSE_2009}
S. Lany and A. Zunger,
Modelling Simul. Mater. Sci. Eng. \textbf{17}, 084002 (2009).

\bibitem{Lambrecht_pssb_2010}
W. R. L. Lambrecht, 
Phys. Status Solidi B, doi:10.1002/pssb.201046327 (2011).

\bibitem{Janotti_APL_2005}
A. Janotti and C. G. Van de Walle,
Appl. Phys. Lett. \textbf{87}, 122102 (2005).

\bibitem{Lany_PRL_2007}
S. Lany and A. Zunger,
Phys. Rev. Lett. \textbf{98}, 045501 (2007).

\bibitem{Janotti_PRB_2007}
A. Janotti and C. G. Van de Walle,
Phys. Rev. B \textbf{76}, 165202 (2007).

\bibitem{Paudel_PRB_2008}
T. R. Paudel and W. R. L. Lambrecht,
Phys. Rev. B \textbf{77}, 205202 (2008).

\bibitem{Pemmaraju_PRB_2008}
C. D. Pemmaraju, R. Hanafin, T. Archer, H. B. Braun, and S. Sanvito,
Phys. Rev. B \textbf{78}, 054428 (2008).

\bibitem{Deak_JPCM_2005}
P. De\'{a}k, A. Gali, A. Solyom, A. Buruzs, and Th.  Frauenheim,
J. Phys.: Condens. Mat. \textbf{17}, S2141 (2005).

\bibitem{Knaup_PRB_2005}
J. M. Knaup, P. De\'{a}k, Th. Frauenheim, A. Gali, Z. Hajnal, and W. J. Choyke,
Phys. Rev. B \textbf{72}, 115323 (2005).

\bibitem{Gavartin_PRB_2006}
J. L. Gavartin, D. Mu\~{n}oz Ramo, A. L. Shluger, G. Bersuker, and B. H. Lee,
Appl. Phys. Lett. \textbf{89}, 082908 (2006).

\bibitem{Broqvist_APL_2006}
P. Broqvist and A. Pasquarello,
Appl. Phys. Lett. \textbf{89}, 262904 (2006).

\bibitem{Oba_PRB_2008}
F. Oba, A. Togo, I. Tanaka, J. Paier, and G. Kresse,
Phys. Rev. B \textbf{77}, 245202 (2008).

\bibitem{Janotti_PRB_2010}
A. Janotti, J. B. Varley, P. Rinke, N. Umezawa, G. Kresse, and C. G. Van de Walle,
Phys. Rev. B \textbf{81}, 085212 (2010).

\bibitem{Deak_PRB_2010}
P. De\'{a}k, B. Aradi, Th. Frauenheim, E. Janz\'{e}n, and A. Gali,
Phys. Rev. B \textbf{81}, 153203 (2010).

\bibitem{Broqvist_pssa_2010}
P. Broqvist, A. Alkauskas, and A. Pasquarello,
Phys. Status Solidi A \textbf{207}, 270 (2010).

\bibitem{Clark_PRB_2010}
S. J. Clark, J. Robertson, S. Lany, and A. Zunger,
Phys. Rev. B \textbf{81}, 115311 (2010).

\bibitem{Segev_PRB_2007}
D. Segev, A. Janotti, and C. G. Van de Walle,
Phys. Rev. B \textbf{75}, 035201 (2007).

\bibitem{Persson_PRB_2005}
C. Persson, Y.-J. Zhao, S. Lany, and A. Zunger,
Phys. Rev. B \textbf{72}, 035211 (2005).

\bibitem{Hedstrom_PRL_2006}
M. Hedstr\"{o}m, A. Schindlmayr, G. Schwarz, M. Scheffler,
Phys. Rev. Lett. \textbf{97}, 226401 (2006).

\bibitem{Rinke_PRL_2009}
P. Rinke, A. Janotti, M. Scheffler, and C. G. Van de Walle,
Phys. Rev. Lett. \textbf{102}, 026402 (2009).

\bibitem{Lany_PRB_2010a}
S. Lany and A. Zunger,
Phys. Rev. B \textbf{81}, 113201 (2010).

\bibitem{Bockstedte_PRL_2010}
M. Bockstedte, A. Marini, O. Pankratov, and A. Rubio,
Phys. Rev. Lett. \textbf{105}, 026401 (2010).

\bibitem{Bruneval_PRB_2011}
F. Bruneval and G. Roma,
Phys. Rev. B \textbf{83}, 144116 (2011).

\bibitem{Giantomassi_pssb_2011}
M. Giantomassi, M. Stankovski, R. Shaltaf, M. Gr\"{u}ning, 
F. Bruneval, P. Rinke, and G.-M. Rignanese,
Phys. Status Solidi B \textbf{248}, 275 (2011).

\bibitem{Kohan_PRB_2000}
A. F. Kohan, G. Ceder, D. Morgan, and C. G. Van de Walle,
Phys. Rev. B \textbf{61}, 15019 (2000).

\bibitem{Erhardt_PRB_2006}
P. Erhart, K. Albe, and A. Klein,
Phys. Rev. B \textbf{73}, 205203 (2006).

\bibitem{Agoston_PRL_2009}
P. \'{A}goston, K. Albe, R. M. Nieminen, and M. J. Puska,
Phys. Rev. Lett. \textbf{103}, 245501 (2009).

\bibitem{Gallino_JCP_2010}
F. Gallino, G. Pacchioni, and C. Di Valentin,
J. Chem. Phys. \textbf{133}, 144512 (2010).

\bibitem{Boonchun_pssb_2011}
A. Boonchun and W. R. L. Lambrecht,
Phys. Status Solidi B \textbf{248}, 1043 (2011).

\bibitem{Perdew_PRL_1996}
J. P. Perdew, K. Burke, and M. Ernzerhof,
Phys. Rev. Lett. {\bf 77}, 3865 (1996).

\bibitem{Becke_JCP_1993}
A. D. Becke,
J. Chem. Phys. {\bf 98}, 1372 (1993);
J. Chem. Phys. {\bf 98}, 5648 (1993).

\bibitem{Perdew_JCP_1996}
J. P. Perdew, K. Burke and M. Ernzerhof,
J.\ Chem.\ Phys.\ {\bf 105}, 9982 (1996).

\bibitem{Alkauskas_PRL_2008b}
A. Alkauskas, P. Broqvist, F. Devynck, and A. Pasquarello,
Phys. Rev. Lett. \textbf{101}, 106802 (2008).

\bibitem{Shimazaki_JCP_2009}
T. Shimazaki and Y. Asai,
J. Chem. Phys. \textbf{130}, 164702 (2009).

\bibitem{Alkauskas_pssb_2011}
A. Alkauskas, P. Broqvist, and A. Pasquarello,
Phys. Status Solidi B \textbf{248}, 775 (2011).

\bibitem{Marques_2010}
M. A. L. Marques, J. Vidal, M. J. T. Oliveira, L. Reining, and S. Botti,
Phys. Rev. B \textbf{83}, 035119 (2011).

\bibitem{Carvalho_PRB_2009}
A. Carvalho, A. Alkauskas, A. Pasquarello, A. K. Tagantsev, and N. Setter,
Phys. Rev. B \textbf{80}, 195205 (2009).

\bibitem{Troullier_PRB_1991}
N. Troullier and J. L. Martins,
Phys. Rev. B {\bf 43}, 1993 (1991).

\bibitem{CPMD}
R. Car and M. Parrinello,
Phys. Rev. Lett. \textbf{55}, 2471 (1985);
CPMD, Copyright IBM Corp 1990-2006,
Copyright MPI f\"{u}r Festk\"{o}rperforschung Stuttgart 1997-2001.

\bibitem{Hutter_CPC_2005}
J. Hutter and A. Curioni,
ChemPhysChem \textbf{6}, 1788 (2005).

\bibitem{Todorova_JPCB_2006}
T. Todorova, A. P. Seitsonen, J. Hutter, I. F. W. Kuo, and C. J. Mundy,
J. Phys. Chem. B \textbf{110}, 3685 (2006).

\bibitem{Broqvist_PRB_2009}
P. Broqvist, A. Alkauskas, and A. Pasquarello,
Phys. Rev. B \textbf{80}, 085114 (2009).

\bibitem{Makov_PRB_1995}
G. Makov and M. C. Payne,
Phys. Rev. B \textbf{51}, 4014 (1995).


\bibitem{Heyd_JCP_2003}
J. Heyd, G. E. Scuseria, and M. Ernzerhof,
J. Chem. Phys. \textbf{118}, 8207 (2003).

\bibitem{Komsa_PRB_2010}
H.-P. Komsa, P. Broqvist, and A. Pasquarello,
Phys. Rev. B \textbf{81}, 205118 (2010).

\bibitem{comment_Lany}
In Ref.\ \onlinecite{Lany_PRL_2007}, LDA+$U_{d}$ was only used 
to determine the shifts of the bulk band edges.
For atomically localized defects this is a valid procedure 
according to Ref.\ \onlinecite{Alkauskas_PRL_2008a}.

\bibitem{LDA-GGA}
In the case of ZnO LDA yields a band gap of about 0.6 eV, which is yet smaller that the one
in the GGA calculation. However, for the sake of simplicity we assume that bulk band edges
are described similarly in LDA and GGA. 
The error this assumption introduces is of the order of 0.1 eV (Ref.\ \onlinecite{Lany_PRB_2008})
and thus unsubstantial for our consideration.


\bibitem{Lyons_PRB_2009}
J. L. Lyons, A. Janotti, and C. G. Van de Walle,
Phys. Rev. B \textbf{80}, 205113 (2009).

\bibitem{Broqvist_APL_2009}
P. Broqvist, J. F. Binder, and A. Pasquarello,
Appl. Phys. Lett. \textbf{94}, 141911 (2009).

\bibitem{Broqvist_APL_2008}
P. Broqvist, A. Alkauskas, and A. Pasquarello,
Appl.\ Phys.\ Lett.\ {\bf 92}, 132911 (2008).

\bibitem{Janak_PRB_1978}
J. F. Janak,
Phys. Rev. B \textbf{18}, 7165 (1978).

\bibitem{Ihm_JPC_1979}
J. Ihm, A. Zunger, and M. L. Cohen,
J. Phys. C: Solid State Phys. \textbf{12}, 4409 (1979)

\bibitem{Perdew_PRL_1982}
J. P. Perdew, R. G. Parr, M. Levy, and J. L. Balduz,
Phys. Rev. Lett. \textbf{49}, 1691 (1982).

\bibitem{Perdew_PRL_1983}
J. P. Perdew and M. Levy,
Phys. Rev. Lett. \textbf{51}, 1884 (1983).

\bibitem{Foster_PRB_2001}
A. S. Foster, V. B. Sulimov, F. Lopez Gejo, A. L. Shluger, and R. M. Nieminen,
Phys. Rev. B \textbf{64}, 224108 (2001).

\bibitem{Pruneda_PRB_2005}
J. M. Pruneda and E. Artacho, 
Phys. Rev. B \textbf{71}, 094113 (2005)

\bibitem{Curtiss_JCP_1997}
L. A. Curtiss, P. C. Redfern, K. Raghavachari, and  J. A. Pople,
J. Chem. Phys. \textbf{109}, 42 (1998).

\bibitem{Heyd_JCP_2004} 
J. Heyd and G. E. Scuseria,
J. Chem. Phys. \textbf{120}, 7274 (2004).

\bibitem{note_pentacene}
The calculations of the pentacene molecule were performed with cubic supercells of
increasing side (30 \AA, 40 \AA, and 50 \AA) and extrapolated to infinity.  
Single particle eigenvalues were aligned to the vacuum level far from the molecule. 
For charged systems, this vacuum level was preserved at the same energy and the 
total energies were consistently corrected.  In addition, the electrostatic 
monopole-monopole correction reflecting the interaction of an array of charges 
with the neutralizing background was included.

\bibitem{Clark_HCA_1972}
P. A. Clark, F. Brogli, and E. Heilbronner,
Helv. Chim. Acta \textbf{55}, 1415 (1972).

\bibitem{Crocker_JACS_1993}
L. Crocker, T. Wang, and P. Kebarle,
J. Am. Chem. Soc. \textbf{115}, 7818 (1993).

\bibitem{Deleuze_JCP_2003}
M. S. Deleuze, L. Claes, E. S. Kryachko, and J. P. Fran\c{c}ois,
J. Chem. Phys. \textbf{119}, 3106 (2003).

\bibitem{Slater}
J. C. Slater,
Adv. Quantum Chem. \textbf{6}, 1 (1972).

\bibitem{Vydrov_JCP_2007}
O. A. Vydrov, G. E. Scuseria, and J. P. Perdew,
J. Chem. Phys. \textbf{126}, 154109 (2007).

\bibitem{Perdew_LNP_2003}
J. P. Perdew and S. Kurth,
Lecture Notes in Physics \textbf{620}, 1 (2003)

\bibitem{Perdew_PRB_1997}
J. P. Perdew and M. Levy,
Phys. Rev. B \textbf{56}, 16021 (1997).


\bibitem{Mori-Sanchez_PRL_2008}
P. Mori-S\'{a}nchez, A. J. Cohen, and W. Yang,
Phys. Rev. Lett. \textbf{100}, 146401 (2008).

\bibitem{Cohen_PRB_2008}
A. J. Cohen, P. Mori-S\'{a}nchez, and W. Yang,
Phys. Rev. B \textbf{77}, 115123 (2008).

\bibitem{Lany_PRB_2009}
S. Lany and A. Zunger,
Phys. Rev. B \textbf{80}, 085202 (2009).

\bibitem{Lany_PRB_2010b}
S. Lany and A. Zunger,
Phys. Rev. B \textbf{81}, 205209 (2010).

\bibitem{Dabo_PRB_2010}
I. Dabo, A. Ferretti, N. Poilvert, Y. Li, N. Marzari, and M. Cococcioni, 
Phys. Rev. B \textbf{82}, 115121 (2010).

\bibitem{Ogut_PRL_1998}
S. \"{O}\v{g}\"{u}t, J. R. Chelikowsky, and S. G. Louie,
Phys. Rev. Lett. \textbf{80}, 3162 (1998);
R. W. Godby and I. D. White,
\emph{ibid.} \textbf{80}, 3161 (1998).

\bibitem{Alkauskas_PB_2007}
A. Alkauskas and A. Pasquarello,
Physica B \textbf{401-402}, 670 (2007).

\bibitem{Chen_PRB_2010}
W. Chen, C. Tegenkamp, H. Pfn\"ur, and T. Bredow,
Phys.\ Rev.\ B {\bf 82}, 104106 (2010).

\bibitem{Komsa_APL_2010}
H.-P. Komsa and A. Pasquarello,
Appl.\ Phys.\ Lett.\ {\bf 97}, 191901 (2010).

\bibitem{Komsa_PRB_2011}
H.-P. Komsa and A. Pasquarello,
Phys.\ Rev.\ B, in print, article ID: BDR1188B.

\bibitem{VanDeWalle_PRB_1987}
C. G. Van de Walle and R. M. Martin,
Phys. Rev. B \textbf{35}, 8154 (1987).

\bibitem{Shaltaf_PRL_2008}
R. Shaltaf, G.-M. Rignanese, X. Gonze, F. Giustino, and A. Pasquarello,
Phys.\ Rev.\ Lett.\ {\bf 100}, 186401 (2008).

\bibitem{Gruning_PRB_2010}
M. Gr\"{u}ning, R. Shaltaf, and G.-M. Rignanese,
Phys. Rev. B \textbf{81}, 035330 (2010).

\bibitem{Wadehra_APL_2010}
A. Wadehra, J. W. Nicklas, J. W. Wilkins, 
Appl. Phys. Lett. \textbf{97}, 092119 (2010).

\bibitem{Devynck_PRB_2007}
F. Devynck, F. Giustino, P. Broqvist, and A. Pasquarello,
Phys. Rev. B {\bf 76}, 075351 (2007).

\bibitem{Shaltaf_PRB_2009}
R. Shaltaf, T. Rangel, M. Gr\"{u}ning, X. Gonze, G.-M. Rignanese, and D. R. Hamann,
Phys. Rev. B \textbf{79}, 195101 (2009).

\bibitem{Shishkin_PRL_2007}
M. Shishkin, M. Marsman, and G. Kresse,
Phys. Rev. Lett. \textbf{99}, 246403 (2007).

\bibitem{Shishkin_PRB_2007}
M. Shishkin and G. Kresse,
Phys. Rev. B \textbf{75}, 235102 (2007).


















\end{thebibliography}
\end{document}